\documentclass[twocolumn,showpacs,preprintnumbers,amsmath,amssymb]{revtex4}
\usepackage{graphicx}% Include figure files
\usepackage{dcolumn}% Align table columns on decimal point
\usepackage{bm}% bold math
\usepackage{epsfig}
\usepackage{amssymb}
\usepackage{psfrag}
\usepackage{natbib}
\usepackage{epsfig}
\usepackage{epstopdf}

\def\bea{\begin{eqnarray}}
\def\eea{\end{eqnarray}}
\def\beq{\begin{equation}}
\def\eeq{\end{equation}}

\def\p#1{\partial_{#1}}

\begin{document}

%\preprint{APS/123-QED}

\title{A nonlinear equation for ionic diffusion in a strong
           binary electrolyte}% Force line breaks with \\

\author{Sandip Ghosal}
 %\altaffiliation[Also at ]{Physics Department, XYZ University.}%Lines break automatically or can be forced with \\
\author{Zhen Chen}%
% \email{Second.Author@institution.edu}
\affiliation{%
Northwestern University, Department of Mechanical Engineering\\
2145 Sheridan Road, Evanston, IL 60208
%This line break forced with \textbackslash\textbackslash
}%

%\author{Charlie Author}
 %\homepage{http://www.Second.institution.edu/~Charlie.Author}
%\affiliation{
%Second institution and/or address\\
%This line break forced% with \\
%}%

%\date{\today}% It is always \today, today,
             %  but any date may be explicitly specified

\begin{abstract}
The problem of the one dimensional electro-diffusion of ions in a strong binary electrolyte is considered. 
In such a system the solute dissociates completely into 
two species of ions with unlike charges. The mathematical description consists 
of a diffusion equation for each species augmented by transport due to a  self consistent 
electrostatic field determined by the Poisson equation. This mathematical framework also describes 
other important problems in physics such as electron 
and hole diffusion across semi-conductor junctions and the diffusion of ions in plasmas. If concentrations do not 
vary appreciably over distances of the order of the Debye length, the Poisson equation 
can be replaced by the condition of local charge neutrality first introduced by Planck.  It can then be shown that both species 
diffuse at the same rate with a common diffusivity that is intermediate between that  of the slow and fast species (ambipolar diffusion). 
Here we derive a more general theory by exploiting the ratio of Debye length to a characteristic length scale 
as a small asymptotic parameter.  It is shown that the concentration of either  species may be described by a nonlinear 
integro-differential equation which replaces  the classical linear equation for ambipolar diffusion but reduces to it in the appropriate limit.
Through numerical integration of the full set of equations it is shown that this nonlinear equation provides 
a better approximation to the exact solution than the linear equation it replaces.
\end{abstract}

\pacs{82.45.Gj,82.45.-h,82.70.y,47.57.eb,47.57.jd,72.20.Dp}% PACS, the Physics and Astronomy
                             % Classification Scheme.
%\keywords{Suggested keywords}%Use showkeys class option if keyword
                              %display desired
\maketitle

%\section{\label{sec:level1}First-level heading:\protect\\ The line
%break was forced \lowercase{via} \textbackslash\textbackslash}
%\section{Introduction}
The one dimensional electro-diffusion equations describing the evolution of ionic concentrations $n(x,t)$, $p(x,t)$ in a 
fully dissociated binary electrolyte with a 
self consistent electrostatic potential $\phi(x,t)$  may be put in the form~\cite{rubinstein_book}
\bea
\p{t}n-\p{x}(n\p{x}\phi)&=&\p{xx}n \label{eq4n}\\
u^{-1} \p{t}p-z\p{x}(p\p{x}\phi)&=&\p{xx}p \label{eq4p}\\
\p{xx}\phi&=&-n-zp \label{poisson}
\eea
where $u=u_p/u_n$ is the ratio of mobilities of  species p  and n and $z=z_p/z_n$ is the ratio of their valences. We will take n as 
the species with lesser mobility, so that $u \geq 1$.
Since the two species are oppositely charged, $z$ is negative. 
The above equations have been rendered dimensionless by scaling the concentrations by the concentration of n at 
infinitely far points ($n_{*}$)  so that  $n(\infty,t)=1$,  the co-ordinate $x$ by a 
length scale $\lambda_{*}$ related to Debye length, the time by a diffusion time $\lambda_{*}^{2}/D_{n}$  and the potential $\phi$ 
by the thermal energy $(k_{B} T) / (e z_n)$. In the above, $D_{n}$ and $D_{p}$ are the diffusivities of the two species which are related to the mobilities $u_{n},u_{p}$
 through the Einstein relation $D_n/u_n = D_p/u_p = k_B T$, where $k_B$ is the Boltzmann constant and 
$T$ the absolute temperature. The length scale $\lambda_{*}$ referred to above is defined by the 
expression $\lambda_{*}^{2} = (\epsilon_{0} \kappa k_{B} T ) / (z_n^{2} e^{2} n_{*} )$
in terms of the electronic charge ($e$), permittivity of vacuum ($\epsilon_0$) and 
dielectric constant ($\kappa$). It is related to the Debye length in a homogeneous 
charge neutral electrolyte where the less mobile species has a  dimensionless concentration of  $n$ as
$\lambda_{D} (n) =\lambda_{*}/\sqrt{n(1-z)}$.

Eq.(\ref{eq4n})-(\ref{poisson}) and its modifications also describe other transport problems of interest in various areas of physics. For example, if 
$\phi (x,t) = - E_{0} x + \phi^{\prime}$ where $E_0$ is an external field and $\phi^{\prime}$ is the perturbation in the potential 
and if a term $-N(x)$ representing the density of free charges is  added to the right hand side of Eq.~(\ref{poisson}), then these equations become the 
Van Roosbroeck equations describing the migration of charge carriers in solid state physics, where $n$, $p$ and $N$ represent the concentrations of electrons, holes and dopants~\cite{markowich_book}. 
If $N(x)$ is interpreted as the density of ion exchange sites within a membrane, then the same set of equations also describes various filtration 
processes such as electrodialysis~\cite{rubinstein_book}. In plasma physics~\cite{krall_trivelpiece_book} 
 one often encounters situations where the electrons and 
positive ions may  be regarded as two interpenetrating charged fluids of different diffusivities
coupled by a self-consistent electric field. It is then described mathematically by Eq.~(\ref{eq4n})-(\ref{poisson}). The generalization 
of these equations to three or more species in the presence of an applied electric field $E_{0}$  describe the separation of 
charged macromolecules in capillary electrophoresis~\cite{ghosal_annrev06} and other separation processes based on electric charge. 

One of the simplest experimental realizations of Eq.~(\ref{eq4n})-(\ref{poisson}) is the ``Liquid Junction'' problem, where a barrier initially separates 
two semi-infinite regions of a binary electrolyte (such as sodium chloride in water) of different concentrations,  and subsequently, at 
time $t=0$, the barrier is removed. Since generally the positive and negative components would diffuse at different rates 
a polarization vector and consequently a measurable ``Liquid Junction Potential'' (LJP) appears across the interface. 
 Planck~\cite{Planck} derived an expression for the LJP on the basis of the local charge 
neutrality  assumption, an idea that has since become a central paradigm in many problems involving electro-diffusion. The physical basis 
of the approximation is that the electrostatic force that would arise if a charge separation was realized is so strong that in practice it precludes 
any departure from electro-neutrality anywhere in space. 

Mathematically, the local electro-neutrality assumption 
amounts to neglecting the term $\p{xx}\phi$ on the left  hand side of Eq.~(\ref{poisson}),
so that it reduces to $p=-n/z$. This relation can now be used to eliminate the terms involving $\phi$ in Eq.~(\ref{eq4n}) and (\ref{eq4p}). 
It then follows that 
$n$ (as well as $p=-n/z$) obeys the diffusion equation 
\beq
\p{t}n = D\p{xx}n \label{ambipolar_diff_eq}
\eeq
where $D=u(1-z)/(1-uz)$, a result due to Henderson~\cite{hendersonA,hendersonB}.
 Since $z<0$, clearly $u \geq D \geq 1$.
The coulomb attraction between the two species enhances the rate of spreading of the slower species and reduces that of the faster species so that 
both diffuse at an equal, ambipolar (that is, the same for either charge) rate. In the analysis outlined above, though the term in $\phi$ is dropped in 
Eq.~(\ref{poisson}), it must be retained in Eq.~(\ref{eq4n})-(\ref{eq4p}), a situation that appeared self contradictory and led to some controversy until
 Hickman~\cite{hickman,jackson} provided a proper interpretation within the framework of an 
asymptotic theory based on expansion in the small parameter $k \lambda_{D}$, where $k$ is a characteristic wave number and $\lambda_D$ is 
the Debye length. Since the Debye length $\lambda_{D}$ is normally a very small quantity (about $3$ nm in a $0.01$ M solution of a common salt like sodium 
or potassium chloride) in most applications the charge neutrality assumption is very accurate. However,  it could be violated in many modern 
applications such as in nanochannels where one or more geometric dimensions may be of the order of nanometers. Such departures from charge 
neutrality may give rise to new effects. It would therefore seem worthwhile to first study these effects 
 in the context of the simple model system represented by Eq.~(\ref{eq4n})-(\ref{poisson}).

In this paper we use the method of multiple scales~\cite{nayfeh_bk} to reduce Eq.~(\ref{eq4n})-(\ref{poisson}) to a nonlinear one dimensional 
system in the limit of long (compared to the Debye length)
length scales  and slow (compared to the diffusion time over a Debye length) time scales. We show that at the
lowest order the equation for ambipolar diffusion is recovered. If the asymptotic theory is continued to the next order, a nonlinear integro-differential 
equation for the concentration $n$ emerges. Numerical solutions of this reduced equation is seen to agree better with that of the full system 
of Eq.~(\ref{eq4n})-(\ref{poisson}) and lead to  effects not captured in the lowest order theory based on local charge neutrality. 

We would like to study the behavior of Eq.~(\ref{eq4n})-(\ref{poisson}) at long length and time scales, under the boundary conditions that $n,p,\phi$ 
all approach constant values and respect local charge neutrality  as $x \rightarrow \pm \infty$. Thus, we are considering passive diffusion 
in the absence of imposed electric fields and currents. 
 Following the usual procedure~\cite{nayfeh_bk}, 
we introduce slow variables $\xi = \sqrt{\epsilon} \, x$ and $\tau = \epsilon \, t$ and suppose that $n,p,\phi$ depend solely on $\xi$ and $\tau$.
Then Eq.~(\ref{eq4n})-(\ref{poisson}) reduce to 
\bea
\p{\tau}n-\p{\xi}(n\p{\xi}\phi)&=&\p{\xi \xi}n \label{eq4n1}\\
u^{-1} \p{\tau}p-z\p{\xi}(p\p{\xi}\phi)&=&\p{\xi \xi}p \label{eq4p1}\\
\epsilon \p{\xi \xi }\phi&=&-n-zp \label{poisson1}
\eea
where $\epsilon$ is a small parameter. Expanding all dependent variables in asymptotic series in $\epsilon$, such as 
$n=n_{0} + \epsilon n_{1} + \epsilon^{2} n_{2} + \cdots$, substituting in Eq.~(\ref{eq4n1})-(\ref{poisson1}) and 
equating the coefficients of $\epsilon$ on both sides, we get a series of equations starting with the lowest order ones: 
\bea
\p{\tau}n_0 -\p{\xi}(n_0 \p{\xi}\phi_0)&=&\p{\xi \xi}n_0 \label{eq4nzero}\\
u^{-1} \p{\tau}p_0 -z\p{\xi}(p_0 \p{\xi}\phi_0)&=&\p{\xi \xi}p_0 \label{eq4pzero}\\
-n_0-zp_0&=& 0 \label{poissonzero}
\eea
If the last equation is used to eliminate the second terms from the first two, we get 
\bea 
\p{\tau}n_0&=&D \p{\xi \xi} n_0 \label{ambipolar}\\
p_0 &=& - n_0 / z \label{LCN}
\eea 
with $D=u(1-z)/(1-uz)$, representing ambipolar diffusion and if  the time derivative 
terms are eliminated instead and the result integrated, we get
\beq 
\phi_{0} = \frac{u-1}{1-uz} \ln n_0 \label{phi0}.
\eeq
If  the above equation is evaluated at  $x=-\infty$, we get Planck's formula~\cite{Planck}
for the potential drop across a liquid junction. At the next order in $\epsilon$, we have 
\bea
\p{\tau}n_1-\p{\xi}(n_1 \p{\xi}\phi_0)-\p{\xi}(n_0 \p{\xi}\phi_1)&=&\p{\xi \xi}n_1 \quad\quad \label{eq4norder1}\\
u^{-1} \p{\tau}p_1-z\p{\xi}(p_1 \p{\xi}\phi_{0})-z\p{\xi}(p_0 \p{\xi}\phi_{1})&=&\p{\xi \xi}p_1  \label{eq4porder1}\\
\p{\xi \xi }\phi_0 = -n_1 -z p_1 &&\label{poissonorder1}
\eea
If we add Eq.~(\ref{eq4norder1}) and (\ref{eq4porder1})  and use Eq.(\ref{poissonzero}), 
the term in $\phi_1$ is eliminated. We then obtain after substituting  $\phi_0$ from Eq.~(\ref{phi0}) 
and after some algebra;
\bea
\p{\tau}n_1&=& D \p{\xi \xi} n_1 +  \alpha  \p{\xi \xi \xi \xi}( \ln n_0 )  - \beta \p{\xi \xi} (\p{\xi}\ln n_0)^{2} \quad\quad  \label{eq4norder1alt}\\
p_1 &=&  - \frac{n_1}{z} - \frac{u-1}{z(1-uz)} \p{\xi \xi} ( \ln n_0 ), \label{depLCN}
\eea
where $\alpha$ and $\beta$ are positive constants defined by 
\bea 
\alpha &=& - \frac{uz(u-1)^{2}}{(1-uz)^{3}} \label{definealpha}\\
\beta &=& \frac{u(u-1) (2-z-uz)}{2(1-uz)^{3}}
\eea 
If we add Eq.~(\ref{ambipolar}) to $\epsilon$ times Eq.~(\ref{eq4norder1alt}) and transform back to our 
original variables $x$ and $t$, we get 
\beq
\p{t}n =  D \p{xx} n +  \alpha  \p{xxxx}( \ln n ) 
  - \beta \p{xx} (\p{x} \ln n)^{2}   \quad \label{evolven}
\eeq
with an error which is higher than order $\epsilon^{2}$.
In arriving at Eq.~(\ref{evolven}) we have replaced the terms 
$\alpha  \p{\xi \xi \xi \xi}( \ln n_0 )$ by $\alpha  \p{\xi \xi \xi \xi}( \ln n )$ 
and $- \beta \p{\xi \xi} (\p{\xi}\ln n_0)^{2}$ by  $- \beta \p{\xi \xi} (\p{\xi}\ln n)^{2}$,
which is justified because doing so introduces an error in Eq.~(\ref{evolven}) 
that is higher  than order $\epsilon^{2}$. Similarly, combining Eq.~(\ref{depLCN}) 
with Eq.~(\ref{LCN}) we get 
\beq
p  =  - \frac{n}{z} - \frac{u-1}{z(1-uz)} \p{xx} ( \ln n )  \label{slavep}
\eeq 
with an error of  order $\epsilon^{2}$. The last term in Eq.(\ref{slavep}) is due to 
the  departure from charge neutrality. 

Let us now consider some of the consequences of Eq.~(\ref{evolven}) and (\ref{slavep}).
First, we  consider weak perturbations from the constant values at infinity: $n=1+n^{\prime}$ 
and $p = -1/z + p^{\prime}$, where $|n^{\prime}|, |p^{\prime}| \ll 1$. This gives a hyper-diffusion 
equation for $n^{\prime}$: 
\beq 
\p{t} n^{\prime} = D \p{xx} n^{\prime} + \alpha \p{xxxx} n^{\prime}.
\eeq 
Therefore, solutions of the form $\exp[ at + ik x ]$ have  growth rates $a = - k^{2} D + k^{4} \alpha$. 
Since $\alpha >0$, high wavenumber modes $k > k_{\text{max}} = \sqrt{(D/\alpha)}$ are unstable. 
We will show that this instability is entirely spurious. Indeed, Eq.~(\ref{evolven}) is not even valid for modes $k > k_{{\text max}} \sim 1$ 
since these solutions violate  the premise $k \ll 1$. The instability arises as a consequence of truncating the asymptotic series,
 as can be seen by considering the linearized  version of Eq.~(\ref{eq4n})-(\ref{poisson}) with $u^{-1}=0$:
\bea
\p{t}n^{\prime} -\p{xx} \phi &=&\p{xx}n^{\prime}, \label{linearizedn}\\
\p{xx} \phi &=&\p{xx}p^{\prime}, \label{linearizedp}\\
\p{xx}\phi&=&-n^{\prime} -zp^{\prime}.  \label{linearizedpoisson}
\eea
If Eq.~(\ref{linearizedp}) is integrated and the result substituted in Eq.~(\ref{linearizedpoisson}) 
an exact expression for $\phi$ may be found in terms of $n^{\prime}$:
\bea 
\phi &=& -\frac{1}{z} (1 + z^{-1} \p{xx} )^{-1} n^{\prime}, \label{phi1}\\
&= & -  \int_{-\infty}^{+\infty} G(|x-y|; \sqrt{-z}) n^{\prime}(y) \, dy, \label{phi2}\\
&=& -\frac{1}{z} \left[ 1- z^{-1} \p{xx} +z^{-2} \p{xxxx} + \cdots \right] n^{\prime} \label{phi3}
\eea 
where $G(x;m)= -  \exp( - m |x| ) / (2m)$ is the Green's function of the Helmholtz operator $\p{xx} - m^{2}$.
If the exact inversion, that is  Eq.~(\ref{phi2}),  is substituted in Eq.~(\ref{linearizedn}) one gets an 
integro-differential equation which in Fourier-space yields a growth rate 
\beq 
a = - k^{2} \left[ 1 + \frac{1}{k^{2} -z} \right] \sim - k^{2} \left( 1 - \frac{1}{z} \right) + \frac{k^{4}}{z^{2}} + \cdots \label{aseries}
\eeq 
The exact formula for $a$ indicated by the $=$ sign above shows that there is no high wavenumber instability if the exact inversion,  Eq.~(\ref{phi2}),
is employed. The second part, indicated by the $\sim$ sign,   shows the result of the approximate inversion, Eq.(\ref{phi3}),  based on the low wavenumber 
approximation.  In this case there is a high wavenumber 
instability if the asymptotic series is truncated after an even number of terms. That is, the root cause of the instability is the 
oscillatory approach to the limit of the series indicated on the right hand side of Eq.~(\ref{aseries}). 

\begin{figure}[t]
%\begin{center}
   \includegraphics[angle=0,width=3.0truein]{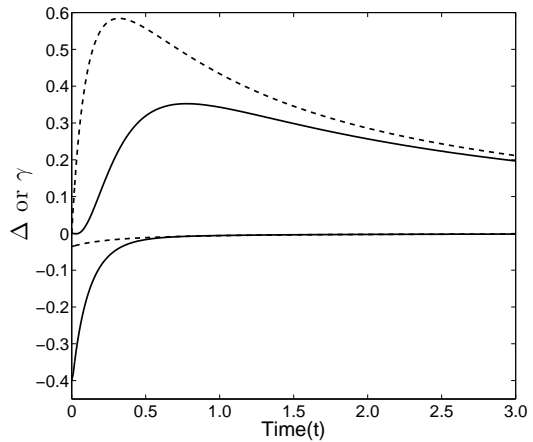}
       \caption{
       The quantity $\Delta = (2D)^{-1} d\sigma^{2}/dt - 1$ (lower quadrants) where $\sigma^{2}$ is the variance of the distribution $n(x,t)$, 
       and the excess kurtosis $\gamma$ (upper quadrants) 
       determined using the models (II) [dashed] and (III) [solid]. For model (I), $\gamma = \Delta = 0$ at all $t$.
        The initial state is  $n(x,0)=1+A \exp(-x^{2}/2)$ and  parameters are  $A=1.0$, $u=6$, $z=-1$.
       }
    \label{figure1}
%\end{center}
\end{figure}
The above analysis suggests a simple way of modifying Eq.~(\ref{evolven}) without violating its asymptotic validity,
 but at the same time eliminating the spurious high wavenumber instability. We recognize 
that the term $\alpha \p{xxxx} (\ln n) = \p{xx} ( \alpha \p{xx} \ln n)$ in Eq.~(\ref{evolven}) most likely resulted from 
a truncated expansion of the Helmholtz operator: $\p{xx} [1- \alpha \p{xx} ]^{-1}  \sim \p{xx} (1+ \alpha \p{xx}  + \cdots ) = \p{xx} + \alpha \p{xxxx} + \cdots$ 
and simply replace $\alpha \p{xx}$ in Eq.~(\ref{evolven}) by $(1- \alpha \p{xx})^{-1} - 1$. Thus, we get the nonlinear 
integro-differential equation: 
\beq
\p{t} n = \p{xx} {\cal F}[n] \label{evolven1}
\eeq 
where ${\cal F}$ is the functional:
\bea
{\cal F}[n] &=& D n  - \beta [ \p{x} (\ln n) ]^{2} \nonumber \\
&+&   \frac{1}{2 \sqrt{\alpha}}\int_{-\infty}^{+\infty} e^{- |x-y| / \sqrt{\alpha} } \ln \left\{ \frac{n(y)}{n(x)} \right\}  \, dy. 
 \label{functional_F}
\eea
Eq.~(\ref{evolven1}) is asymptotically equivalent to Eq.~(\ref{evolven}) since they differ only by terms of order higher than $\epsilon^{2}$ 
but it is free from the spurious high wavenumber instability. By virtue of its construction it also has the property that when $u^{-1}=0$, the linearized version of Eq.(\ref{evolven1}) is the exact solution of Eq.(\ref{linearizedn})-(\ref{linearizedpoisson}). Thus, when $u$ is large and amplitudes are 
small, the solutions of Eq.(\ref{evolven1}) match closely the true solution, irrespective of the validity of the assumption of long length scales and slow time scales
exploited in the asymptotic theory.
The formal condition for the validity of Eq.~(\ref{evolven1}) is $\lambda_{D} \bar{n}^{-1} \p{\bar{x}} \bar{n} \ll  1$, 
where the bars on top of the variables indicate that they are in dimensional form. In dimensionless notation, the condition of 
validity reduces to $|\p{x} n|/ [n^{3/2} \sqrt{1-z}] \ll 1$. It should be noted that, Eq.~(\ref{evolven1}) as well as the lower order 
effective diffusivity approximation fails  in the limit of very low concentrations ($n \rightarrow 0$). This is because the 
Debye length becomes infinitely large in this limit invalidating our premise of long length scales compared to the local 
value of the Debye length. Eq.~(\ref{evolven1}) together with Eq.~(\ref{slavep}) to calculate the concentration of the faster 
diffusing species are our principal results. 

A numerical method employing a fourth order finite difference algorithm for spatial derivatives coupled with a fourth order Runge-Kutta time stepping scheme with  fixed grid and step size
was used to solve a time dependent electrodiffusion problem. An initial condition $n(x,0)=1+A \exp(-x^{2}/2)$ was used with $A=1.0$ and the other parameters were set as $u=6$ and $z=-1$. 
The problem was solved at three levels of approximation: \\[1ex]
(I) Using the ambipolar diffusion equation, Eq.(\ref{ambipolar_diff_eq}).\\
(II) Using the nonlinear equation, Eq.(\ref{evolven1}).\\
(III) Using the full electro-diffusion model  Eq.~(\ref{eq4n})-(\ref{poisson}).\\[1ex]
\noindent It is clear that with (I) the variance $\sigma^{2}$ would increase linearly with time, but not so in the case of (II).
Further,  a distribution that is initially 
Gaussian remains so at future times under (I)  but not under (II).
Therefore, the excess kurtosis $\gamma$ is expected to remain zero at all times in the case of (I) but not in the case of (II). 
In Figure~1 the quantities 
$\Delta \equiv (2D)^{-1} d \sigma^{2} / dt - 1$ and $\gamma$ are plotted as a function of time ($t$) for  (II) 
and (III). In case of  (I), $\Delta = \gamma  = 0$ at all $t$ and this case is not shown for clarity. 
It is seen that at sufficiently long times (I), (II) and (III) all approach a common asymptotic limit. 
However, at shorter times, (II) is in better accord with (III) than (I) is. 

The qualitative nature of the time variation of $\gamma$ and $\Delta$ may also be understood on the basis 
of Eq.~(\ref{evolven1}). To see this, we use  Eq.(\ref{phi3}) to expand the integral operator and put Eq.~(\ref{evolven1}) 
 in the form
\beq
\p{t}n =  D \p{xx} n +  \alpha  \p{xxxx}( \ln n ) 
  - \beta \p{xx} (\p{x} \ln n)^{2}   + \cdots \quad \label{evolven2}
\eeq
where $\cdots$ indicate terms involving sixth, eighth, tenth, .... order derivatives. If we further restrict ourselves to small amplitudes,
 $n = 1 + n^{\prime}$, with $|n^{\prime}| \ll 1$, then $\ln n = n^{\prime} - \frac{1}{2} (n^{\prime})^{2} + \cdots$, so that the Eq.(\ref{evolven2})  becomes 
\beq
\p{t}n^{\prime}  =  D \p{xx} n^{\prime}  + \alpha  \p{xxxx} n^{\prime} -  \frac{\alpha}{2}   \p{xxxx} (n^{\prime})^{2}
  - \beta \p{xx} (\p{x} n^{\prime} )^{2}   + \cdots  \label{evolven3}
\eeq
Here the $\cdots$ now include the nonlinear terms of higher order. If we multiply Eq.~(\ref{evolven3}) by $x^{k}$ and 
integrate, we can generate ordinary differential equations for the moments $m_k  =  \int_{-\infty}^{\infty} x^{k} n^{\prime} (x) \; dx$, starting with $dm_{0}/dt=dm_{1}/dt=0$.
The neglected sixth and higher derivative terms  do not contribute for $k \leq 4$. Without loss of generality, we can assume $m_{1}=0$ so 
that here $\sigma^{2}= m_{2}/m_{0}$ and $\gamma=m_{4} m_{0} / m_{2}^{2} - 3$. By combining the equations for the second and 
fourth moments, we derive
\beq 
\frac{d \sigma^{2}}{dt} = 2 D - \frac{2 \beta}{m_{0}} \int_{-\infty}^{+\infty} ( \p{x} n^{\prime} )^{2} \; dx \label{eq4var}
\eeq 
and 
\beq 
\frac{d \gamma}{dt} = -  \frac{4 \gamma D}{\sigma^{2}}  + \frac{12 \alpha}{\sigma^{4}} + \cdots \label{eq4kurt}
\eeq 
where $\cdots$ indicate the nonlinear terms that we suppress. Now, if $\alpha=0$,  then either $\gamma=0$ (if it is zero initially) 
or else $\gamma$ monotonically decays to zero. When deviations from the ambipolar diffusion limit is considered ($\alpha > 0$), 
we find that $\gamma$ initially increases (when $\sigma$ is small enough for the second term to dominate), 
but as $\sigma$ becomes larger, the first term eventually dominates resulting in a decrease in $\gamma$ 
towards zero. Further, Eq.(\ref{eq4var}) shows, that since $\beta > 0$ the rate of increase of the variance is slightly 
less than  $2D$ but the  deficit in the growth 
rate becomes progressively smaller at large times. This is indeed what is observed in Figure~1.

In summary, a prototypical problem in electro-diffusion was considered in the limit where the Debye-length was non-zero 
but nevertheless small compared to a characteristic scale of spatial variation. It was shown that in this limit the concentration 
can be described by a one dimensional nonlinear integro-differential equation which reduces to the linear diffusion 
equation describing ambipolar diffusion if all but the leading order terms in the ratio of Debye-length to a characteristic spatial scale are 
neglected. Numerical solution shows that the nonlinear integro-differential equation provides a better approximation 
to the true solution. The approach described here 
could be useful for other problems in electro-diffusion where the Debye length is small but may not be considered negligible 
in relation to other length scales. \\[1ex]
\noindent {\em Acknowledgement:}  Support from the NIH under grant R01EB007596 is gratefully acknowledged.

\bibliographystyle{/Users/sandipghosal/Documents/G5docs/LIBRARY/BIBFILES/prsty}
\bibliography{/Users/sandipghosal/Documents/G5docs/LIBRARY/BIBFILES/membrane,/Users/sandipghosal/Documents/G5docs/LIBRARY/BIBFILES/microfluidics}
%\bibliography{../BIBFILES/membrane,../BIBFILES/microfluidics}

\end{document}